%% file: paper.tex
\documentclass[letterpaper,twocolumn,10pt]{article}
\usepackage{usenix2019_v3}

\usepackage{graphicx}
\usepackage{subfig}
\usepackage{listings}
\usepackage{tabularx}
\usepackage{booktabs}
\usepackage{multirow}
\usepackage[acronym]{glossaries}
\usepackage{cleveref}
\usepackage[square,sort,comma,numbers]{natbib}

\usepackage[scaled=0.9]{inconsolata}

\definecolor{discussgreen}{rgb}{0,0.6,0}
\definecolor{todoblue}{rgb}{0,0,0.6}
\newcommand{\discuss}[1]{}
\newcommand{\todo}[1]{}
\newcommand{\hidefornow}[1]{}
\newcommand{\latertodo}[1]{}
\newcommand{\laterdiscuss}[1]{}

\newcommand{\unics}[0]{\textsc{uni-cs}}
\newcommand{\unilaw}[0]{\textsc{uni-law}}
\newcommand{\private}[0]{\textsc{citizen}}

\newcommand{\letter}[0]{\textsc{letter}}
\newcommand{\mail}[0]{\textsc{email}}  

\newcommand{\privacy}[0]{\textsc{privacy}}
\newcommand{\gdpr}[0]{\textsc{gdpr}}
\newcommand{\fee}[0]{\textsc{gdpr+fine}}

\newcommand{\control}[0]{\textsc{control}}

\begin{document}

\pagenumbering{gobble}

\title{Effective Notification Campaigns on the Web:\\A Matter of Trust, Framing, and Support}

  
\author{{\rm Max Maass}\\TU Darmstadt
\and
{\rm Alina Stöver}\\TU Darmstadt
\and
{\rm Henning Pridöhl}\\Universität Bamberg
\and 
{\rm Sebastian Bretthauer}\\Goethe-Universität Frankfurt
\and
{\rm Dominik Herrmann}\\Universität Bamberg
\and
{\rm Matthias Hollick}\\TU Darmstadt
\and
{\rm Indra Spiecker}\\Goethe-Universität Frankfurt
}


\maketitle
\input{sections/99-glossary}

\input{sections/00-abstract}
\input{sections/01-intro}

\input{sections/02-relatedwork}

\input{sections/03-background}
\input{sections/04-methodology}
\input{sections/05-results}
\input{sections/06-survey}
\input{sections/07-discussion}

\input{sections/09-conclusion}
\input{sections/10-acknowledgements}

\clearpage

\bibliography{bibliography}
\bibliographystyle{abbrvnat-pp}

\input{sections/11-appendix}
\end{document}

%% file: sections/99-glossary.tex
\glsdisablehyper
\newacronym{DKIM}{DKIM}{DomainKeys Identified Mail}
\newacronym{SPF}{SPF}{Sender Policy Framework}
\newacronym{TLD}{TLD}{Top-Level Domain}
\newacronym{GDPR}{GDPR}{General Data Protection Regulation}
\newacronym{CERT}{CERT}{Computer Emergency Response Team}
\newacronym{ISP}{ISP}{Internet Service Provider}
\newacronym{GA}{GA}{Google Analytics}

%% file: sections/00-abstract.tex
\begin{abstract}
Misconfigurations and outdated software are a major cause of compromised websites and data leaks.
Past research has proposed and evaluated sending automated security notifications to the operators of misconfigured websites, but encountered issues with reachability, mistrust, and a perceived lack of importance\laterdiscuss{ wir wollten weg von importance---"pressure to act" passt hier aber auch nicht so richtig gut. Ideen?}.
In this paper, we seek to understand the determinants\todo{ we moved away from that term everywhere else, do we want to keep it around here or change it?} of effective notifications.
We identify a data protection misconfiguration that affects 12.7\,\% of the 1.3 million websites we scanned and opens them up to legal liability.
Using a subset of 4754 websites\todo{with manually collected address information?}, we conduct a multivariate randomized controlled notification experiment, evaluating contact medium, sender, and framing of the message.
We also include a link to a public web-based self-service tool that is run by us in disguise and conduct an anonymous survey of the notified website owners (N=477) to understand their perspective.

We find that framing a misconfiguration as a problem of legal compliance can increase remediation rates, especially when the notification is sent as a letter from a legal research group, achieving remediation rates of 76.3\,\% compared to 33.9\,\% for emails sent by computer science researchers warning about a privacy issue.
Across all groups, 56.6\,\% of notified owners remediated the issue, compared to 9.2\,\% in the control group.
In conclusion, we present factors that lead website owners to trust a notification, show what framing of the notification brings them into action, and how they can be supported in remediating the issue.

\end{abstract}

%% file: sections/01-intro.tex
\section{Introduction}
\todo{sicherstellen, dass keine todos und discuss von leerzeichen umschlossen sind, sonst hat man beim auskommentieren der todos am ende doppelte Leerzeichen im PDF}

Maintaining a website has become a complex endeavor that requires keeping software up to date and adapting configurations to changing technical requirements.
It is thus inevitable that some systems will not be updated in time, leading to vulnerabilities and data breaches like the Equifax breach, traced back to a missing software update \cite{ArsEquifax}, or the Exactis leak, which was caused by a misconfigured ElasticSearch instance \cite{WiredExactis}.
Such breaches routinely violate the privacy of millions of people and can cost companies millions of dollars in remediation costs, settlements, and regulatory fines \cite{Enforcementtracker}.


Past research has evaluated the possibility of sending automated notifications to system operators to inform them about insecure \cite{Durumeric2014,Li2016WWW,Stock2016,Stock2018}, compromised \cite{Vasek2012,Canali2013,Cetin2016,Cetin2018,Cetin2019}, or misconfigured \cite{Kuhrer2014,Li2016Usenix,Cetin2017,Zeng2019} systems under their control.
Such attempts achieved an improvement in remediation compared to a control group, but often found large numbers of systems to remain unfixed.


These studies found varied and in some cases contradictory results on the determinants of successful notifications.
We seek to shed light on factors that influence the success of a notification.
In particular, we consider the following research questions:
(1) What influence do various factors of notifications, such as the contact medium, the sender, and the framing of the problem, have on remediation?
(2) Which forms of support are desired and embraced by website owners?
(3) What properties of a notification message lead site owners to trust or distrust it?

As the subject of our notifications, we search for a misconfiguration that (a) results in non-compliance with legal obligations, (b) exposes website owners to an immediate financial risk, (c) can be automatically and unambiguously detected upon visiting a site, and (d) is straightforward to fix.
These requirements are met when website owners use the third-party service \emph{\gls{GA}} in Germany but fail to turn on the \emph{IP Anonymization} feature.
Both Google and the supervisory authorities place the responsibility of enabling IP Anonymization with the site owners.
A German court has recently convicted a site owner, who failed to enable this feature, on grounds of violating personal privacy rights \cite{LGDresden}.
Choosing a compliance misconfiguration instead of a security vulnerability ensures that the notification is equally relevant to all website owners.
Unlike outdated software, which sometimes cannot be updated due to compatibility issues, there is also no incentive \emph{not} to remediate.

Our scans identified this misconfiguration on 12.7\,\% of the approximately 1.3 million German websites we analyzed.
We conduct a notification experiment with a subset of 4594 site owners operating 4754 distinct non-compliant websites, for which we collect contact information \emph{manually}.
We send notifications via email or letter, using three different senders and three distinct framings. 
The notification contains a link to a public self-service tool that we run in disguise to enable site owners to verify if the problem has been resolved.
We also support owners via phone and email.
Finally, after two months and one reminder, we send a debriefing message and invite all notified owners to answer a short survey to gain an understanding of their perception of the notification message.

In summary, our paper makes the following contributions:
\begin{itemize}
    \item We scan for a common misconfiguration, which can be framed as a compliance issue as described in \autoref{sec:background}.
    This misconfiguration allows us to design a covert randomized controlled notification experiment to evaluate the effect of three factors on remediation in \autoref{sec:methodology}.
    \item We report on the results of our notifications in \autoref{sec:results}, finding high remediation rates between 33.9 and 76.3\,\%, with the control group at 9.2\,\%.
    We observed a high demand for the support mechanisms we provide, in particular for our self-service tool.
    \item We describe the responses collected in our survey ($N=477$) in \autoref{sec:survey}, finding that missing awareness is widespread.
    19.5\,\% of site owners admitted not even knowing that their site was running \gls{GA}. 
    \item We highlight important takeaways from our study, in particular the large effect of framing misconfigurations as an issue of compliance with legal obligations in \autoref{sec:discussion}.
\end{itemize}

%% file: sections/02-relatedwork.tex
\section{Related Work}
\label{sec:relatedwork}
We review previous research in the area of vulnerability notifications and the perspectives of system operators and owners.

\subsection{Effectiveness of Notifications}
\label{sec:effectiveness}
The effectiveness of notifications was evaluated in several areas, ranging from the security of websites \cite{Vasek2012,Canali2013,Durumeric2014,Cetin2016,Li2016WWW,Stock2016,Stock2018,Zeng2019} or DNS servers \cite{Cetin2017} to DDoS amplifiers \cite{Kuhrer2014,Li2016Usenix,Cetin2019} and end-user malware infections \cite{Cetin2018}, with studies usually finding an increase in remediation rates compared to a control group.
The studies commonly sent emails to WHOIS or abuse contacts, or to common aliases (RFC 2142), with some using
intermediaries such as CERTs and clearinghouses \cite{Kuhrer2014,Li2016Usenix,Stock2016,Cetin2017}.
Some studies also worked with Google \cite{Li2016WWW,Zeng2019,Li2016Usenix} or \glspl{ISP} using quarantine networks with captive portals \cite{Cetin2016,Cetin2019} to deliver their messages.
Stock \emph{et al.} performed a smaller-scale experiment with manually collected email addresses, postal addresses, phone numbers, and social media contacts \cite{Stock2018}, finding that these channels can sometimes outperform others.
However, the low number of messages ($N=364$ spread over 10 groups) and potential priming and self-selection effects in this experiment make it impossible to draw general conclusions.

Studies frequently encountered issues with notification delivery \cite{Stock2016,Stock2018,Cetin2016,Durumeric2014,Li2016Usenix,Cetin2017}, observing email bounce rates of over 50\,\% \cite{Cetin2017,Stock2016,Cetin2016} in some cases due to incorrect information in WHOIS records or the lack of standard email aliases such as \emph{webmaster@domain.com}.
Additional issues in delivery were caused by spam filters \cite{Stock2016,Stock2018}.
Recipients were often wary of unsolicited emails and suspected them to be spam or scam messages \cite{Stock2018,Cetin2018,Cetin2019,Zeng2019}, sometimes reaching out to verify the veracity of the message before acting upon it \cite{Cetin2018,Cetin2019}.
This suggests that trust in the sender could play an important role in the success of notifications.
However, other studies did not find significant differences between different senders \cite{Cetin2016,Zeng2019}, leaving this question unresolved.

Several studies showed that more comprehensive messages increase remediation rates \cite{Vasek2012,Li2016Usenix,Cetin2016} and trust in the message~\cite{Stock2018}. 
Recipients often desired a tool to verify the veracity of the provided information and effectiveness of their remediation \cite{Zeng2019,Cetin2017,Li2016WWW}, although the actual effect of providing such a tool was small \cite{Cetin2017}.
The results for repeated notifications of unfixed websites are inconclusive as well, showing no effects in a study conducted by Li \emph{et al.} \cite{Li2016Usenix}, while Stock \emph{et al.} observed a small effect \cite{Stock2016}.

An area where notices are arguably followed \emph{too} well is in copyright enforcement, where the financial risks surrounding the \emph{notice and takedown} scheme have led to 
overblocking and incorrect claims \cite{Perel2017,Urban2017}.
This highlights the potential impact the legal incentives surrounding a notification can have.

In contrast to existing work, we seek to study alternative delivery channels and senders in more detail and investigate the effect of using a compliance argument to provide an incentive for remediation that is independent of circumstances, as it applies equally to every website.

\subsection{Website Owners Perspective}
With our notification, we explicitly addressed the owners in contrast to previous work that wrote to operators, as they are legally responsible for the operation of the website and thus the correct point of contact for a compliance issue.
The owner can also be the one to operate the system, which is why in the following, we also refer to the literature on \emph{system operators}.
While there is quite a lot of research regarding the developers' perspective on privacy (e.\,g. \cite{ayalon2017, van_der_linden_data_nodate, senarath2018, hadar2018}), and some about system operators \cite{alt2019, dietrich2018}, relatively little research into website owners exists so far \cite{ginosar2017}.
Similar to software developers \cite{xu2013}, system operators play a critical role when it comes to protecting end-users privacy and security \cite{alt2019}.
This may also be true for website owners since they are making decisions regarding specific privacy policies and their implementation \cite{ginosar2017}.

Research about whether or not system operators and website owners are aware of security and privacy gaps is ambiguous.
Many consumer and small business site owners are not aware that their sites are threatened \cite{Commtouch}.
In the notification study by Li \emph{et al.} \cite{Li2016Usenix}, 46\,\% of the participants stated that they were aware of their vulnerability before notification, while Durumeric \emph{et al.} \cite{Durumeric2014} reported that all 17 participants were already aware of the problem.
Even if the operators were aware, they did not necessarily solve the problem in these studies, which was also concluded by Stock \emph{et al.} \cite{Stock2018}. 



This brings us to the question of how system operators or website owners handle security and privacy issues.
Many studies pick one specific aspect, e.g., the usability of HTTPS deployment \cite{krombholz2017},
operator's procedures for handling software updates \cite{crameri2007,jenkins2020,li2019},
their perception of the trustworthiness of TLS certificates \cite{ukrop2019},
and their perspectives on TLS misconfigurations \cite{dietrich2018}.
Indications on the system operators' problem-solving behavior are shown by Li \emph{et al.} \cite{li2019}, who describe system operators' processes for software updates with a five-step model.
Dietrich \emph{et al.} \cite{dietrich2018} conducted a study on the perspective of system operators on misconfigurations and found that social, structural, and institutional factors, in particular, can promote a bad security posture \cite{dietrich2018}.

To deepen the understanding of website owners’ perspective on privacy,
we investigate the owners’ reaction to notifications that address a privacy issue.
We focus on the website owners’ awareness, their perception of the notification, and their problem-solving behavior, as well as support aspects.

%% file: sections/03-background.tex
\section{Background}
\label{sec:background}
\glsreset{GA} 
We explain the technical and legal aspects of IP Anonymization in \gls{GA} in Germany, which provides the basis for our notification experiment.
\subsection{Technical Background}
\gls{GA} uses a JavaScript library that has to be included in the website by the website’s owner. 
The owner creates one or more tracker objects with their tracking IDs and adds a method call that 
issues an HTTP request to Google’s Analytics service.
Optionally, the owner can set configuration options on the tracker objects, including IP Anonymization \cite{GoogleAIP}.
When enabling IP Anonymization, the HTTP request contains a parameter \texttt{aip=1}, which instructs Google to truncate the website users’ IP address before storing it for analytical purposes.
For IPv4 addresses, the last octet is set to zero, while for IPv6 addresses, the last 80 bits are changed into zeros.
Configuring IP Anonymization in \gls{GA} is error-prone (see \hyperref[app:misconfiguration]{Appendix A3}).


The effect of IP Anonymization on real-world privacy is limited.
However, choosing this issue for our study has three benefits: it is under the exclusive control of the website operator, can be irrefutably detected remotely, and forms a data protection law violation, which we will discuss next.

\subsection{Legal Background}
The requirement to use IP Anonymization results from the European \gls{GDPR} \cite{GDPR}, and has recently been confirmed in a German sub-court decision \cite{LGDresden}.
The court ruled that omitting IP Anonymization infringes on the data protection principles of data minimization and storage limitation as well as the non-use of pseudonymization and anonymization techniques. 
Enforcement usually falls to the data protection supervisory authorities, which share this interpretation of the law.
While it has not yet been confirmed by a higher court, it at least indicates that non-compliant website operators are at risk of a lawsuit.

German competition law also allows for competitors of the owners of a non-compliant website to send a written warning with costs (``Abmahnung''), a practice that has seen some misuse in the past.
This has led to media attention and fears that the new data protection legislation would result in a wave of such warnings.
While no large number of such cases have appeared so far, many website owners are nevertheless aware of the risk and thus especially sensitive to the topic of \gls{GDPR} compliance.
Website owners bear joint responsibility for the data protection practices of any third-party content they load into their website \cite{CJEUFacebook}, thus placing any detected \gls{GA} code within their legal responsibility.
This makes the issue particularly suitable for evaluating the effects of citing legal requirements in notifications.

Another aspect of the German legal system makes it particularly suitable for notification studies: an imprint with up-to-date contact information is legally mandated for almost all website owners.
While not machine-readable, this improves the chance to identify a point of contact for the website. 

%% file: sections/04-methodology.tex
\section{Methodology}
\label{sec:methodology}
\begin{figure*}
\centering
    \includegraphics[width=1\textwidth]{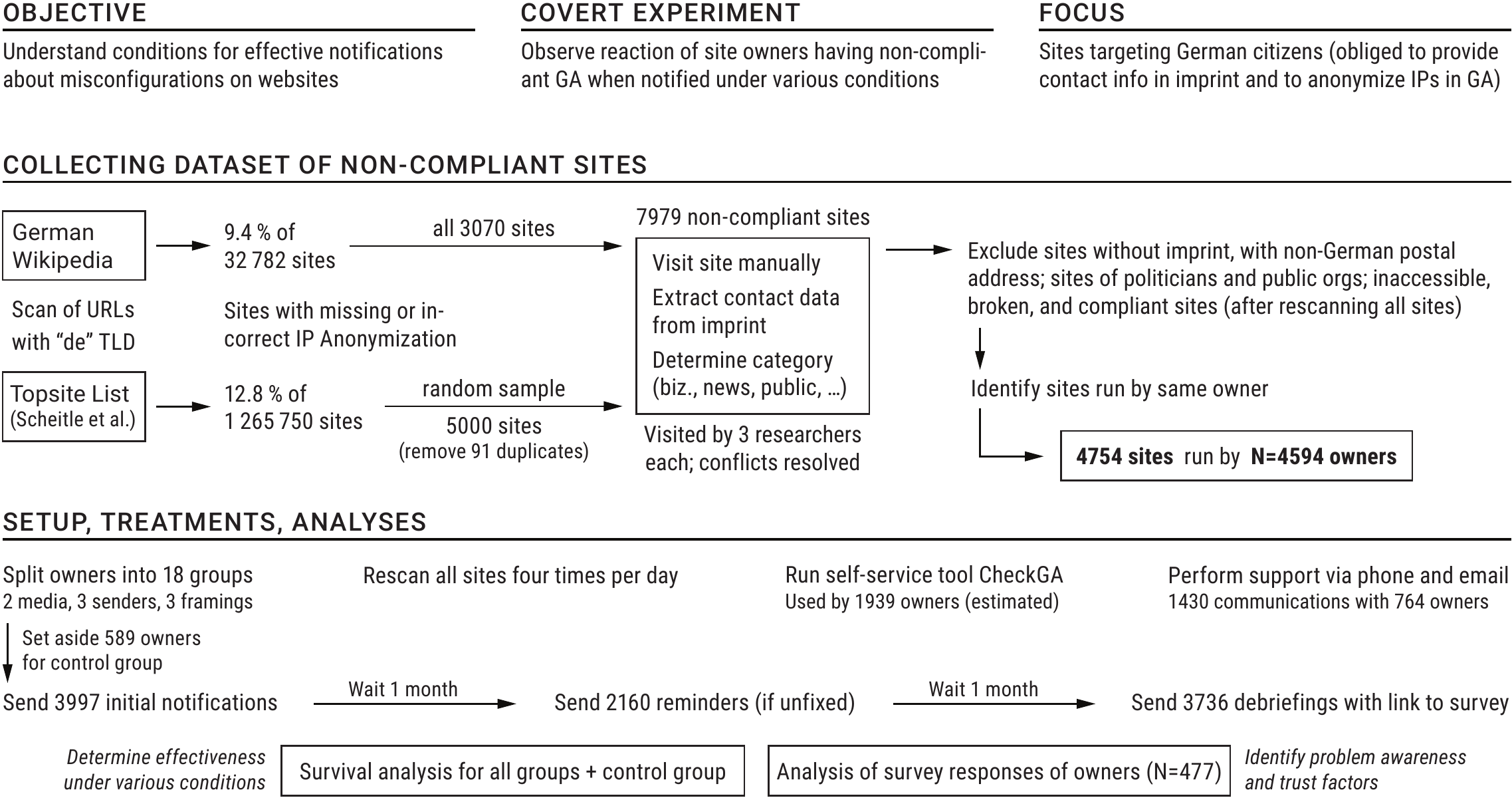}
    \caption{Methodology Overview}%
    \label{fig:methodolgy-overview}%
\end{figure*}

We describe how we collect misconfigured websites, the experimental groups, our notification strategy, and how we supported website owners.
We then present the survey, the steps of data cleaning and analysis, and ethical aspects.
\autoref{fig:methodolgy-overview} summarizes our methodology.

\subsection{Compliance Checker} 
\label{sec:methodology:scanner}
To find misconfigured websites, i.\,e., German sites without IP Anonymization, and to support site owners in verifying the correct implementation of IP Anonymization (cf. \autoref{sec:methodology:tool}), we implement a \emph{compliance checker}.
Our compliance checker is based on the Chromium browser and utilizes the Chrome DevTools protocol \cite{ChromeDevTools}.
It extracts all HTTP requests to GA and checks for the existence of the \texttt{aip=1} parameter, regardless of the request being issued by the website itself or by a third party.
Besides HTTP requests, the checker reads all tracker objects and their configuration on the website in all JavaScript contexts, thus also detecting tracker objects of third parties, e.\,g. included through widgets.
To find tracker objects, the checker iterates all JavaScript global variables.
For each global variable, the checker assesses the available methods and attributes; if those match the expected ones for \texttt{analytics.js} or \texttt{ga.js}, it found a GA object.
This object can be queried for available trackers and their configuration.
Tracker objects are used in our self-service tool to provide the user with detailed information about misconfigurations.

We modify the user agent to hide that Chromium was running headless.
We also scroll the page for a random amount in short random intervals, since additional GA requests might be sent on site usage.
We do not check sites for the presence of consent banners; technically, a consent banner could hide the existence of a non-compliant GA instance until the consent was confirmed, i.\,e., our checker is subject to false negatives.
However, the checker will never return a false positive match.

\subsection{Collecting Non-Compliant Websites}
\label{sec:methodology:addrcollect}

As reported in \autoref{sec:effectiveness}, automated and simple address collection approaches can lead to high bounce rates.
Therefore, we decided to collect all contact data manually.

There is no generally accepted method to obtain a representative sample of websites that fit the criteria for our compliance-based notifications.
Given this limitation, which is discussed in \autoref{sec:limitations},
we still aim to obtain a diverse set of websites for our study, comprised of popular and non-popular sites.
First, we use all websites referenced in the German Wikipedia, filtered on the \gls{TLD} \texttt{.de} ($N$ = 32\,782).
Second, we use a merged and deduplicated version of the archive of historical (up to 10 years) Internet toplists by Scheitle \emph{et al.} \cite{Scheitle2017}, again filtered on the \gls{TLD} \texttt{.de} ($N$ = 1\,265\,750).
We scan these sites with the compliance checker and find that 3070 (9.36\,\%) of the Wikipedia sites and 161\,984 of the toplist sites (12.8\,\%) are non-compliant.
From the non-compliant toplist sites, we randomly sample 5000 sites, 91 of which were already present in the Wikipedia dataset,
to obtain 7979 non-compliant sites in total.

Each site is visited by three researchers, each of them independently collecting postal and email addresses from the site's imprint.
Moreover, the researchers assign a category such as company, individual, public sector, and others, which is used to avoid biases in our experimental groups (cf. \autoref{sec:notific-and-reminder}). Conflicts are discussed and resolved using a majority vote.
On average, this task took 75 seconds per site.

We exclude 3225 sites to which our compliance-based notification does not apply.
About 20\,\% of these sites are excluded because they \emph{belong to the public sector} (municipalities, universities, etc.)
to which the fines mandated by the GDPR do not apply, which would skew some of our results.
We also exclude sites \emph{without an imprint} (about 20\,\% of excluded sites) and sites that \emph{list an address outside of Germany in their imprint} (again, about 20\,\% of excluded sites).
Finally, we exclude sites of politicians (less than 1\,\% of excluded sites)
to avoid cross-contamination with another study.
We also remove sites that cannot be retrieved (about 10\,\% of excluded sites).
Finally, we rescan all sites before sending out the first round of notifications
and remove sites that have become compliant or went out of service during the six-month data collection period
(accounting for the remainder, i.\,e., about 30\,\% of the excluded sites).
Note that the given percentages are only rough estimates as the criteria are not mutually exclusive
and exclusion decisions were often made on the grounds of the most obvious criteria.

As an owner may run more than one website, we merge sites sharing the same owner (\emph{co-owned sites}) into one notification.
To find co-owned sites, we sort postal addresses by ZIP code and manually merge sites with identical or similar addresses (including recipient name).
Addresses are deemed similar if they show only a small variation (e.\,g., ``Company'' vs. ``Company LLC'').
When the recipient or company name differs, sites are not considered co-owned.
We also merge sites with identical contact email addresses.

After merging co-owned sites, we end up with 4754 sites run by 4594 different owners.
These websites were automatically scanned four times a day during the study timeframe.    


\subsection{Notification and Reminder}
\label{sec:notific-and-reminder}
We assign sites randomly to groups defined by three different experimental factors and a control group.
We use a full factorial design (i.\,e., all combinations of factors are used), and assign co-owned sites (cf. \autoref{sec:methodology:addrcollect}) to the same group.
We ensure that the different categories of sites (private, business, etc.) are spread evenly across the groups (stratification).

All messages were sent in German and contain a reference to our self-service tool (cf. \autoref{sec:methodology:tool})\hidefornow{which checks for compliance problems with \gls{GA}}, referencing it as a service that was unaffiliated with the senders of the message, and noting that it may prove helpful in validating the remediation.
Translated versions of our messages can be found in the supplementary material \cite{SupplementaryMaterial}.

\paragraph{Factors}
We differentiate between \textbf{two different contact media}: \letter{} and \mail{}.
As many previous studies have investigated the effect of emails, we choose to emphasize letters in our study by assigning twice the number of websites to the letter groups than the email groups.
Emails are sent in plain-text and contain the entire message in their body (no attachments or external content like tracking pixels).

We compare \textbf{three different senders}: a private individual (\private{}), a computer science group at one university (\unics{}), and a law group at another university (\unilaw{}).
For the two university groups, emails are sent from purpose-specific accounts (\emph{notification@group.university.tld}).
Letters use the official letterheads of the groups, including its return address.
Both emails and letters contain three options for contacting the sender: a postal address, an email, and a phone number.
For the private sender, we use a fresh email account, and letters use the residential address, but no phone number; assuming citizens typically do not provide it.

In the messages, we used \textbf{three different framings} for the problem.
In the \privacy{} framing, we argue that the misconfiguration was harmful to the privacy of website visitors, not mentioning the legal consequences.
In the \gdpr{} framing, we mention that the misconfiguration is violating the \gls{GDPR}.
In the \fee{} framing, we use the same message, but additionally mention the fines that can be leveled against website owners under the \gls{GDPR} (i.\,e., up to 4\,\% of annual turnover). 

\paragraph{Notification, Reminder and Survey}
We sent up to three messages to every recipient: An initial notification, followed by a reminder one month later (if the problem had not been addressed), and a final debriefing message to all contacted recipients a month later to inform them that they had been part of a study, invite them to answer a survey,
and give them the opportunity to opt-out.\discuss{MM: Langer Satz. Schlimm?}
If we received an indication that a message was not deliverable on the selected contact medium (e.\,g., a bounce message from a mail server or our letter being returned), we excluded that recipient from further messages. 

Due to human error, all \unilaw{} – \letter{} reminders were sent with the \fee{} framing.
We will discuss the impact of this mistake in \autoref{sec:results:notify:reminder}.

\subsection{Self-Service Tool and Support}
\label{sec:methodology:tool}
In previous studies, users wished for a self-service tool to check for the reported issue \cite{Zeng2019,Cetin2017,Li2016WWW}.
Besides providing such a tool, we also offer personal support.

\paragraph{Self-Service Tool}
We operate a web-based tool (\emph{CheckGA}) in disguise, i.\,e., not affiliated with \private{}, \unilaw{}, or \unics{}.
Deceiving recipients about the tool's operator increases the trust in our notifications
since recipients can verify the claimed issue with a tool run by an independent, trustworthy organization:
a research group at a German university.

CheckGA allows anyone to analyze the usage of \gls{GA} for arbitrary websites.
The tool has been written for this study and not yet explicitly advertised to others.
However, it is linked from the website of a university chair that researches privacy and security topics, increasing its trustworthiness and making it indexable by search engines.
Some recipients also shared CheckGA on social media and forums.

Users enter a URL to scan, which is then visited by our compliance checker (see \autoref{sec:methodology:scanner}).
The user gets a report about the included GA tracker objects, including trackers added by third parties (e.\,g., due to widgets) and their configuration, such as whether IP Anonymization is enabled.
The report also shows all HTTP requests to the \gls{GA} service, each with the associated analytics data, as well as whether that request contains the \texttt{aip=1} parameter, i.\,e., Google truncates the visitor’s IP address.
The user gets a summary indicating that either everything is fine (no GA found or IP Anonymization correctly implemented) or pointing out the problem.
CheckGA also assists users by providing a help page with extensive documentation, including common pitfalls and code examples.

For each scan, we store the URL of the site, the scan result, the time of the scan, a truncated IP address, and the TLS Session ID.
The last two pieces of data allow us to link scans of different websites conducted by the same user without having to ask users for consent to set cookies.
We informed users before scanning that usage information is collected.


\paragraph{Support via Phone, Email, Letters}
In addition to CheckGA, we also answer phone calls, emails, and occasionally letters from the contacted site owners, assuring them that the messages are authentic, providing basic troubleshooting advice where requested, and addressing complaints by some recipients.
These interactions are described in \autoref{sec:results:complaints}.

\subsection{Survey}
To investigate the site owners' perspective, we invite all contacted owners to participate in a survey after informing them that their sites had been part of a study.
The survey was hosted on the platform \emph{soscisurvey} and consists of an informed consent form and questions regarding the perception of our notification, problem awareness, and solving.
It also contains questions about our and other check tools geared towards system owners, asks if they would like to receive further notifications, and collects basic information about the participants’ affiliation.
The questions are tailored to the group of the participants (medium, sender, framing, final compliance status).
The survey includes between 17 and 21 questions, depending on the group of the recipient and their replies.
A translated version of the survey can be found in the supplementary material \cite{SupplementaryMaterial}.
The responses are analyzed using SPSS. Open answers are analyzed with qualitative content analysis.
The software MAXQDA was used for support. 
561 owners took part in our survey.
We exclude 84 survey answers because the participants either did not agree to the informed consent (N = 19) or answered less than 50\,\% of the questions (N = 65).
226 of the 477 participants completed the questionnaire. 


\subsection{Data Cleaning}
After concluding the data collection, we found that some websites frequently changed the domain they were forwarding to, as they were run by advertising agencies that sold the incoming traffic to different customers over time.
As our scans were based on the URL before following all redirects, we were redirected to different websites and thus do not have continuous reliable data for the domain of the owner we notified.
We thus exclude 31 websites that forwarded to three or more different domains within the study timeframe.

We also found that all sites hosted on the free tier of \emph{Wordpress.com} contained a \gls{GA} tracker managed by Wordpress.
As the owners of the 22 affected sites depended on a centrally-administrated change from \emph{Wordpress.com}\footnote{\emph{Wordpress.com} remediated the misconfiguration after communication with one of the notified website owners.} and none of them contained any additional trackers, we excluded them from the evaluation.
Finally, we remove another two domains from the dataset that were incorrectly labeled as German but were, in fact, run by non-German entities, and four domains from owners that requested to be excluded from the study.

\subsection{Survival Analysis}
\label{sec:methodology:survival}
To evaluate the effectiveness of our notifications, we employ survival analysis.
Previously used by several studies in this field \cite{Vasek2012,Cetin2016,Li2016Usenix,Cetin2018,Zeng2019}, survival analysis operates on data where the event of interest (i.\,e., the website becoming compliant) is still in the future at the time of the analysis (\emph{right-censored} data).
Survival analysis uses estimators like the Kaplan-Meier estimator \cite{Kaplan1958}, which gives us a survival function, i.\,e. a function $S(t)$ that tells us the probability of a misconfiguration surviving past a specific time $t$.
When it comes to notifications, a low survival rate is desirable, as it corresponds to a high remediation rate.

Our evaluation shows that co-owned websites (cf. \autoref{sec:methodology:addrcollect}) tend to show similar remediation behavior.
A more detailed analysis is given in  \hyperref[app:grouping]{Appendix A2}.
To avoid a single owner having a large influence on the results of a group, we compensate for such groupings by using a weighted Kaplan-Meier fit~\cite{pepe1989}.
Each website has a weight associated with it, which is defined as $w=1/|G|$, where $|G|$ denotes the number of websites run by the same owner, thus leading to each owner having the same impact on the results, regardless of the number of websites they operate.
With these weights in place, we ask ``how did our message impact the \emph{owners}'' rather than ``how did our message impact the \emph{websites}.''
We use the \emph{lifelines} library \cite{lifelines} for our analysis.


We run the analysis on the data collected by our automated scanning system that visited every website four times per day.
To avoid transient scan errors impacting the results, we consider a website compliant once $c$ consecutive readings indicate it is \textbf{either not using \gls{GA} or using it with IP Anonymization}, ignoring readings that indicate that the website is offline unless we obtain $c$ consecutive offline readings.
Offline sites will be considered separately.
The different website categories (cf. \autoref{sec:methodology:addrcollect}) show similar behavior, so we do not consider them further in the evaluation.
For our evaluation, we set $c=5$, repeating the evaluation with $c=3$ and $c=8$ and finding equivalent results.
Survival analysis can only work with a \emph{single remediation event} per subject, i.\,e., once a website becomes compliant, the statistics assume it to remain so.
We will check whether this applies in \autoref{sec:results:longterm}.

We cannot use the standard log-rank significance test usually recommended in survival analysis, as our dataset does not fulfill the \emph{proportional hazard} assumption. Instead we compare the functions at specific points in time (before the reminders are sent, and at the end of the study), using a $log(-log(\cdot{}))$-transform, as described by Klein \emph{et al.} \cite{Klein2007}, and the Holm-Bonferroni multi-test correction \cite{Holm1979}.

\subsection{Ethical Considerations}
While our messages are intended to help the recipients avoiding costly mistakes, processing of our messages takes time and, in some cases, money and may cause stress for the owners. 
We consider this risk acceptable, as the message does not contain any demands or threats, and the changes shield the operator from liability.
The contact addresses are collected from the imprint of the website, which is intended for this purpose.

Our scans of the websites only require a normal page load and thus should not put significant strain on their infrastructure.
CheckGA could potentially be used to identify targets for cease-and-desist\latertodo{term} letters.
Since the underlying detection technology could be easily reimplemented by others, we consider the benefits to outweigh the potential harms of providing such a dual-use system.
The data protection compliance of the CheckGA tool was ensured in consultation with legal experts.

While the first two messages do not reveal that they are sent as part of a study to avoid priming effects, we inform all contacted website owners that they were part of a study\hidefornow{ together with the invitation to participate in the survey at the end of the measurement period}.
We respect the wishes of four website owners to be removed from the study.
Members of the control group were informed before this paper was published.
The study was approved by the ethics committee of two of the three involved institutions.
The third institution does not offer a process for ethics approval, but we received approval from the dean of the department.

%% file: sections/05-results.tex
\section{Results}
\label{sec:results}
\begin{figure*}
\centering
    \includegraphics[width=1\linewidth]{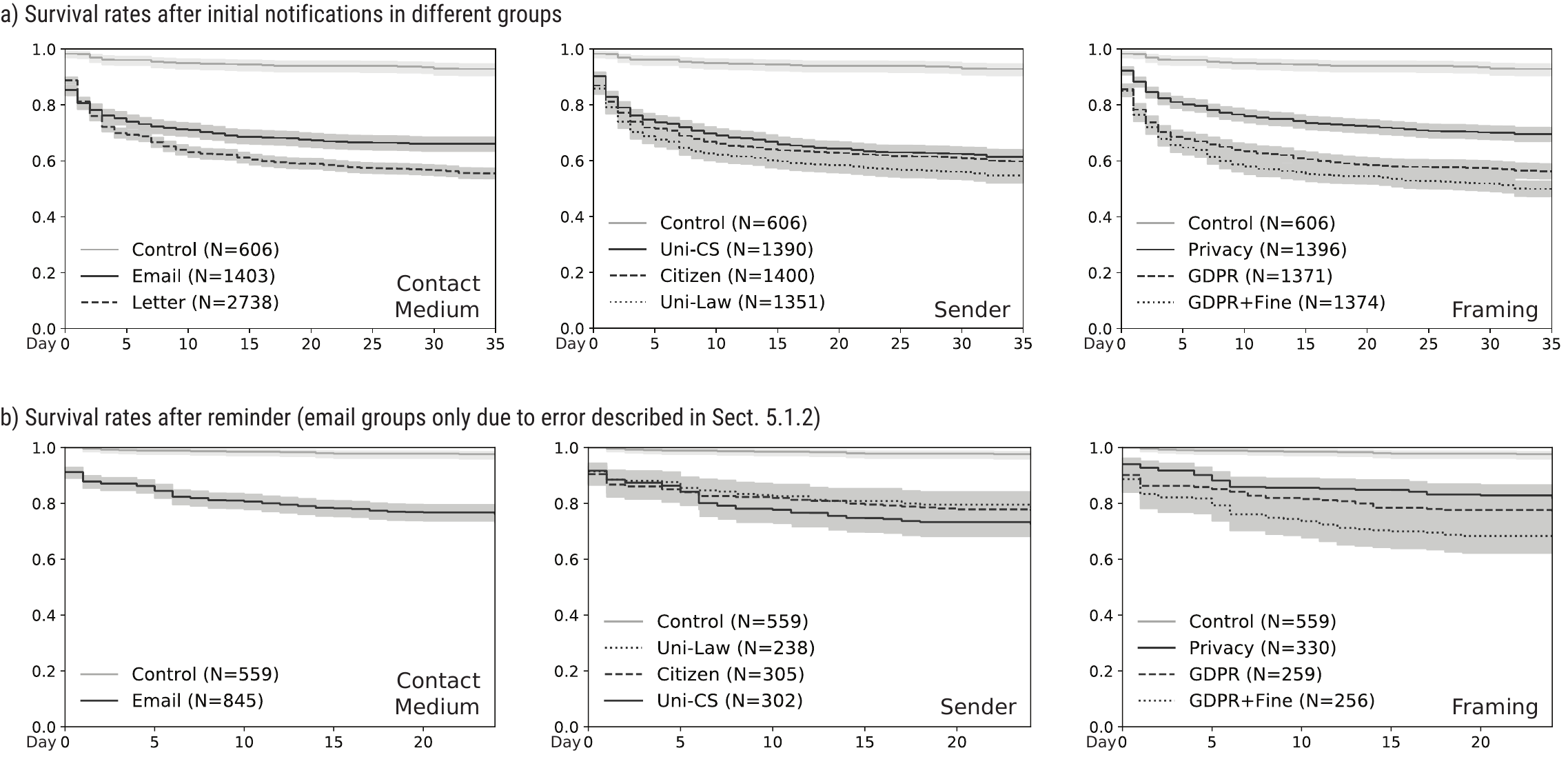}
    \caption{Survival rates after initial notification and reminders \latertodo{MH: kommenar siehe pdf: I do not see an easy way to show the $\pm$ for each curve such that it can still be seen; color?! Error bars?! MM: We're not opening that can of worms again now. }}%
    \label{fig:survival-groups}%
\end{figure*}
We investigate the impact of notifications, the use of CheckGA, our interactions with owners, and long-term effects.
A detailed discussion in relation to the survey results (cf. \autoref{sec:survey}) follows in \autoref{sec:discussion}.

\subsection{Notifications}
\label{sec:results:notify}
We present the 
impact of the notifications on the different groups, using survival analysis (cf. \autoref{sec:methodology:survival}).

\subsubsection{Initial Notification}
\label{sec:results:notify:initial}

The first set of notifications was sent from July 1st to 5th, 2019.
Letters were sent on the Friday of the previous week to ensure they arrived in the same week as the emails, which were spread over five days to avoid triggering rate-based spam filtering.
In total, 48 out of 1337 emails (3.5\,\%) and 153 out of 2660 letters (5.8\,\%) could not be delivered and were returned to the sender.
The number for emails must be considered a lower bound, as many spam filters discard messages silently.

\paragraph{Survival Analysis}
To avoid skewed data due to the staggered sending of notifications, survival times are calculated from the date the message is expected to be received, i.\,e. the day it was sent for emails and July 1st for letters.
The given survival rates and significance values are computed for the last day before the reminders were sent (26–35 days after the initial notifications).
The results of the significance tests are shown in detail in \hyperref[appx:significance]{Appendix A1}.
The given $p$-values are already corrected for multiple comparisons using Holm-Bonferroni \cite{Holm1979} and thus considered significant at $p\leq0.05$.
 
In our survival analysis, all notification groups show an improvement over \control{}.
\hyperref[fig:survival-groups]{Figure~\ref*{fig:survival-groups}a} shows the survival rates (lower is better) for the different varied factors and the confidence interval given by Kaplan-Meier.
For the contact medium, \letter{} had the lowest survival (survival rate $55.6 \pm 1.9\,\%$), significantly lower than the $66.3 \pm 2.6\,\%$ for \mail{} ($p<0.0001$).
For the different senders, the \unilaw{} group led to the most remediations, achieving a survival of $55.0 \pm 2.8\,\%$.
The \private{} group came in second with $59.9 \pm 2.7\,\%$ survival, followed by the \unics{} group with $61.4 \pm 2.7\,\%$.
However, only the difference between \unics{} and \unilaw{} is statistically significant at $p<0.05$.
Finally, for the different framings, the \fee{} framing had the lowest survival ($50.1 \pm 2.8\,\%$), compared to $56.6 \pm 2.8\,\%$ for \gdpr{} and $69.6 \pm 2.6\,\%$ for \privacy{} (all differences were statistically significant).

Comparing the overall highest and lowest survivals of all 18 groups shows the true range of results: while the worst group (\unics{} -- \mail{} -- \privacy{}) resulted in a five-week survival of $82.0 \pm 7.5\,\%$, the best group (\unilaw{} -- \letter{} -- \fee{}) significantly reduced it to $39.4 \pm 5.6\,\%$ (see \autoref{tab:survivalratesappx} in the Appendix), i.\,e., more than 60\,\% of owners remediated the misconfiguration.
This indicates that the considered factors can make a significant difference, although even the worst-performing notification group is still an improvement over sending no notification at all, which shows a survival of $93 \pm 2.4\,\%$ in the same timeframe.
In all cases, the survival curves drop sharply at the beginning.
Most websites are remediated within 7–10 days.

\paragraph{Websites Going Offline}
Some owners took their websites offline instead of remediating the \gls{GA} installation.
In total, 59 non-\control{} websites (1.4\,\%) were offline at the end of the five-week period.
Some owners told us that the websites were outdated and no longer needed.
In the same timeframe, six websites (1\,\%) in \control{} went offline.

\subsubsection{Reminders}
\label{sec:results:notify:reminder}
We sent a reminder to all owners that received our initial message (i.\,e., it did not bounce) but had not become compliant by the 25th of July 2019.
Owners that had contacted us to give updates or ask questions received a hand-crafted reminder, if appropriate.
Email reminders were sent on the 1st and 2nd of August.
For organizational reasons, letters were only sent on the 6th of August.
Even though we did not attempt to contact owners where the delivery of the initial message had failed, five out of 809 reminder emails ($0.6\,\%$) and 27 out of 1351 reminder letters ($2\,\%$) were returned as undeliverable.

As mentioned before, due to human error, we sent the \fee{} framing to all three \letter{} -- \unilaw{} groups, contaminating the results.
However, we present a brief evaluation of the effects of this mistake.

\paragraph{Survival Analysis}
For the post-reminder survival analysis, we only consider owners that received a reminder (i.\,e., we exclude those that had already made their site compliant or where the initial message could not be delivered).
For \control{}, we include sites that were still non-compliant as of the 2nd of August, 2019.
In \hyperref[fig:survival-groups]{Figure~\ref*{fig:survival-groups}b}, we show the post-reminder survival for the different groups, considering only the \mail{} and \control{} due to the unknown influence of the incorrect reminders.
It thus cannot be directly compared with \hyperref[fig:survival-groups]{Figure~\ref*{fig:survival-groups}a}. 

Results for all groups are shown in \autoref{tab:survivalratesappx} in the Appendix\latertodo{MH: maybe make it an own Appendix section: Appendix A2 and force it on the page closeby; currently scattered between other stuff. MM: Hard to do because LaTeX likes to float stuff around}.
\unilaw{} -- \letter{} -- \fee{} achieved a survival of $54.7 \pm 10\,\%$ after 24 days.
Interestingly, the group with the highest survival was also a \unilaw{} group (\unilaw{} -- \mail{} -- \privacy{}), achieving only $88.1 \pm 9.1\,\%$ survival, which is still an improvement over \control{} ($97.6 \pm 1.7\,\%$).
The overall trends remain similar to the initial message, although with smaller differences between the groups.

\paragraph{Accidental Experiment: Increasing the Pressure}
\begin{table}[]
    \centering
    \caption{Survival $S$ in percent and sample size $N$ of \unilaw{} -- \letter{} groups after initial notification ($i$) and reminder ($r$), survival differences to \fee{} in gray.
    Results marked with $\dagger$ erroneously received the \fee{} framing.\laterdiscuss{MH: Make clear that post-reminder is a different dataset. MM: Do we need to re-clarify that here? It's written out in the reminder explanations. AS: würde es so lassen, bin mir nicht sicher, ob weitere Erklärung an der stelle einfach zu viel sind}}
    \begin{tabular}{cllr|lr@{}}
        \toprule
        & Group & $S_i$ & $N_i$ & $S_r$ & $N_r$ \\
        \midrule
        \parbox[t]{2mm}{\multirow{3}{*}{\rotatebox[origin=c]{90}{\letter{}}}} & \fee{} & 39.4 & 304 & 54.7 & 117 \\
        & \gdpr{} & 55.6 {\color{gray} +16.2} & 294 & 68.5 {\color{gray} +13.8} $\dagger$ & 148\\
        & \privacy{} & 62.5 {\color{gray} +23.1} & 293 & 70.4 {\color{gray} +15.7} $\dagger$ & 169 \\
        \bottomrule
    \end{tabular}
    \label{tab:pressure}
\end{table}
The erroneously sent reminders provide us with the opportunity to study the effects of starting with a regular notification and then increasing the pressure with a later letter that explicitly mentions potential fines.
As this experiment was unplanned, we do not have a control group to compare against and thus can only describe the observed values without a baseline for comparison.
However, we can compare it with data from the initial notification.
We thus take a closer look at the results from the \unilaw{} -- \letter{} group, shown in \autoref{tab:pressure}.

Surprisingly, the survival rate for \gdpr{} was 13.8 percentage points higher than that for \fee{}, with \privacy{} showing an even higher survival.
This seems counterintuitive, as one might expect the groups that had previously received a less severe message to be ``shocked into action'' and thus have at least as many remediations as the \fee{} group.

We have no definitive explanation for this behavior.
However, when sending out the survey invitations at the end of the study, we found that some recipients had started recognizing our messages and stopped reading them in detail, with some asking us why we were notifying them again about an issue they had remediated, not realizing that the message contained an invitation to a survey.
Thus, some recipients may have simply recognized the letterhead, remembered the old message, and acted according to that.

\paragraph{Websites Going Offline}
After the reminder, 31 additional websites (including two in \control{}) were offline.

\subsection{CheckGA Usage}
\label{sec:results:tool}
\begin{table}[]
    \caption{Survival rate $S$ and CheckGA usage of all ($U_a$), remediated ($U_r$), and unremediated ($U_u$) owners after initial notification and at the end of the study.}
    \centering
    \begin{tabular}{@{}lcrrrr@{}}
        \toprule
        Group & $S$ [\%]    & ~ ~ & $U_a$ & $U_r$ & $U_u$ \\
        \midrule
        Pre-reminder & 58.8 & & 33.9 & 65.1 & 12.5 \\
        End of study & 43.4 & & 46.9 & 67.6 & 19.8 \\
        \midrule
        \control{} (end of study)  & 90.8 & & 3.1  & 14.8 & 1.9  \\
        \bottomrule
    \end{tabular}
    \label{tab:tooluse}
\end{table}
We now evaluate our web-based tool CheckGA, which site owners used to verify their IP Anonymization.
CheckGA performed 38\,485 scans for 14\,023 sites in total.
12\,047 of the sites are not contained in our dataset.
As we did not advertise the tool, one may assume that those sites that are in our dataset were predominantly scanned by their respective owners.
This assumption is corroborated by the small fraction of scanned sites from \control{} (3.1\,\%).
Under this assumption, half of the notified owners (46.9\,\%) have used the tool at least once for their site(s).
\autoref{tab:tooluse} shows the assumed fraction of owners who used the tool and compares owners who remediated the issue ($U_r$) with those who did not ($U_u$).

\begin{figure}
\includegraphics[width=0.479\textwidth]{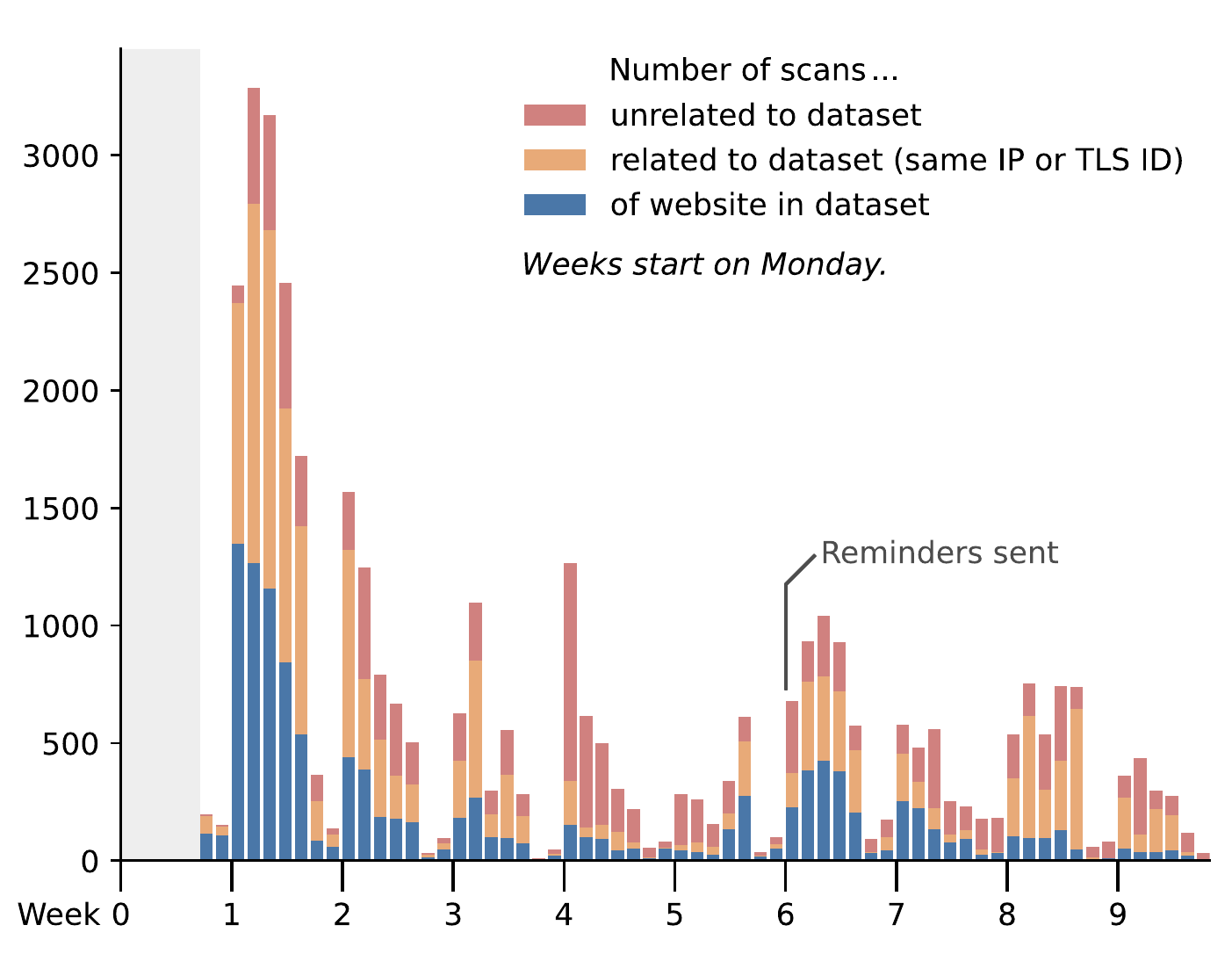}
\caption{User-initiated CheckGA scans per day}
\label{fig:scansperday}
\end{figure}

\paragraph{Scans Over Time}
\autoref{fig:scansperday} shows the number of scans per day during our observation phase of 9 weeks.
First notifications were sent on Friday of Week~0 (cf. \autoref{sec:results:notify}).
A scan is considered a scan of a website in the dataset if either the domain for the user-provided URL is in the dataset itself or redirects to a domain that is in the dataset.
Related scans are those in which likely site owners of our study scan other sites not contained in the dataset.
We define a scan to be related if it targets a site that is not in the dataset, but there is another scan targeting a site in the dataset, and both scans are performed by the same user, identified by the same TLS session or truncated IP address on the same day.
All other scans are considered unrelated to the dataset.

\paragraph{Achieving Compliance}
We also evaluate the number of CheckGA scans performed until a site in our dataset becomes compliant.
For that, we count the scans until \emph{all} user-initiated subsequent scans find the site to be compliant.
In between, a site might appear compliant because the owner rendered GA non-functional while trying to enable IP Anonymization.
Users perform a median of two scans before a site is either remediated or stays non-compliant without further scans, without major differences in mean (4.5 vs. 4.16). Thus, users either get IP Anonymization right quickly or give up early.

It took sites a median of 2.22 hours from the first scan to remediation, with a considerably larger mean of 5.05 \emph{days}.
The fastest 25\,\% of remediating sites became compliant within 3.3 minutes; however, it took over 28 hours to reach 75\,\% compliance, indicating that there are no outliers, but a significant amount of site owners who need an extended time to remediate.
Considering the lower number of scans, site owners who need an extended time possibly reach out for help or pass the issue within their organization.

\subsection{Support and Complaints}
\label{sec:results:complaints}
During the study, we were in contact with many owners who asked questions about our notification, requested help, or questioned the veracity and authenticity of our message.
In total, we received 946 emails (not counting auto-replies), 41 letters, and 56 phone calls from 764 recipients.
We sent 374 emails, one letter, and issued twelve phone calls in reply.

\paragraph{Authenticating the Message}
In total, 32 recipients (4.2\,\% of those in contact with us) contacted us to verify that the message was authentic.
They often chose a different contact address by searching for the sender online and contacting them via their personal addresses listed on the university homepage, or calling phone numbers they found online or in the letter.
Two contacted the sender via Twitter.
The tone of the messages was often friendly and curious, but sometimes hostile, alleging bad intentions or complaining that the message was hard to understand.
Most could be placated with a cover story without mentioning that they were part of a study.

\paragraph{Requesting Help}
204 recipients (26.7\,\%) asked questions about how to remediate the misconfiguration, requested verification of their remediation, or sometimes even offered us login information for the webserver---so we can fix the problem for them, ``if it is that important to you''.
We provided instructions on addressing the misconfiguration but did not take any actions to remediate the websites directly.

\paragraph{Complaints}
19 recipients (2.5\,\%) complained about our messages.
While some were simply unhappy with the unsolicited message or expressed that the tone of the message had been stressful for them, others went further and threatened legal action, tried to bill us for the time they spent on our notification, or even contacted the chancellor of one involved university to complain directly.
We placated these recipients and removed them from future messages upon request.
The assistance of our legal collaborators proved invaluable in many cases.
No legal action was filed against the involved researchers or universities.

\paragraph{Thanks}
Finally, we also received messages of gratitude from 260 recipients (34\,\%), ranging from simple messages to offers of payment, discounts, or gifts. Some recipients sent unsolicited packages with gifts, ranging from free magazines and mugs to a donation to one involved university. Whenever possible, we turned down any offered gifts or payments.

\subsection{Repair vs. Removal}
\label{sec:results:repairremove}
So far, we have treated \gls{GA} being anonymized and completely removed from a website as equivalent (cf. \autoref{sec:methodology:survival}).
However, for site owners, this difference is important, as it changes the insight they get into the behavior of their users.
Surprisingly, we found that of the notified owners that became compliant, $36\,\%$ did so by completely removing \gls{GA} from their site.
This behavior was largely consistent across all experimental groups, indicating that it was not related to any specifics of the notification.
To investigate the correctness of this result, we visited 50 of these pages and manually confirmed that they had removed Google Analytics (and not simply hidden it behind a cookie consent banner), finding no false negatives.

\subsection{Long-Term Effects}
\label{sec:results:longterm}
Our analysis so far only considered whether the problem was solved, but not if it stayed solved.
To answer this question, we crawled all 4754 websites in the study again at the beginning of April 2020 (7 months after the end of the study) to evaluate how many of the previously-compliant websites had become non-compliant again.
Out of 2224 websites that had become compliant at the end of the study period, 78 ($3.5\,\%$) were non-compliant in April (6 of the 78 in \control{}).
Another 38 ($1.7\,\%$) were unreachable.
We thus see a long-term effectiveness of approximately $95\,\%$.


Conversely, of the 2371 sites (550 of which in \control{}) that remained non-compliant at the end of the study period, 438 non-control (24.1\,\%) and 82 from the control group (14.9\,\%) were compliant by the beginning of April (not checking consent banners).
Another 63 were unreachable.
Thus, the base rate of remediations is low (14.9\,\% over 7 months), and the notifications seem to have caused a slight increase in the remediation rate even after the study.

%% file: sections/06-survey.tex
\section{Survey}
\label{sec:survey}
To understand their perspective, we invited the website owners to participate in a survey in the debriefing message.
The survey is shown in the supplementary material \cite{SupplementaryMaterial}.
Responses from 477 owners are included in the following analysis.
The value of participants $N$ varies because the survey did not include any obligatory questions and some items were follow-up questions or only shown for certain groups.


\subsection{Problem Awareness}
371 out of 461 (80.5\,\%) website owners knew that they were using GA on their website before being notified.
272 out of 462 (58.9\,\%) had heard of the IP Anonymization feature before being notified.
58 out of 458 (12.7\,\%) were aware of the missing IP Anonymization before being notified.
We asked those website owners whose IP Anonymization had not been remediated why the problem has not been solved yet ($N=54$; multiple responses possible).
22 owners responded that the problem was unknown, 20 responded that they did not know how to solve the problem.
Some owners mentioned that the problem had no priority (12 responses), they did not find time to deal with the issue (10 responses), or the notification did not seem serious (6 responses).

\subsection{Trust in Notification}
\label{sec:results:perception}

\begin{figure}
\centering
    \includegraphics[width=1\linewidth]{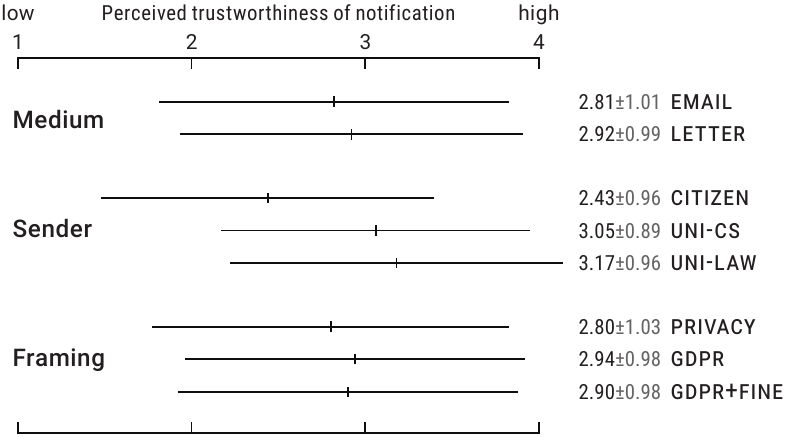}%
    \caption{Agreement of website owners with the statement that the notification made a trustworthy impression.}%
    \label{fig:likert}%
\end{figure}

In the survey 316 out of 460 (68.7\,\%) website owners (rather) agreed with the statement that the notification made a trustworthy impression.
The notification from the law group was perceived most trustworthy and the one from the citizen least trustworthy.
For the remaining two factors, the differences are less pronounced (cf. \autoref{fig:likert}).

Which factors led participants to trust or distrust our message?
To investigate this, we asked \hidefornow{them }two open questions, with 377 participants responding to the trust question and 252 participants to the distrust question (multiple responses possible).
The resulting trust-related factors can be grouped into formal, content-related, and verifiability aspects.

\paragraph{Formal Factors affecting Trust}
Among the formal factors, the sender appears to be of particular importance, being named 348 times in total (multiple responses possible).
Especially the reference to the university, which was mentioned by 174 out of 377 (46.1\,\%) participants, seems to be relevant.
The possibility to contact the sender is also important, being mentioned by 44 out of 377 participants (11.7\,\%).
Another formal aspect is the good use of language mentioned by 63  out of 377 (16.1\,\%) participants, e.\,g., that the notification was ``well-formulated'' and ``did not contain spelling mistakes''.
Of the 259 respondents that had received a letter, 25 (9.6\,\%) named the fact that it was ``a real letter'' as trust-promoting.
Interestingly, even small aspects like the logo or letterhead (13.0\,\%; 49 out of 377 respondents) and signature (3.1\,\%; 12 out of 377 respondents) were named as trust-promoting by some, illustrating that even seemingly small design decisions can have an impact on the perceived trustworthiness.
However, the same factors were also causing mistrust for other recipients, with 41 participants out of 252 (16.3\,\%) mentioning bad wording and 46 (18.2\,\%) the layout as leading to distrust the notification.
12 participants (4.8\,\%) \hidefornow{even }stated that receiving a \emph{letter} was decreasing trust, with one participant wondering, ``Why would anyone even bother to send a letter?''

\paragraph{Content-Related Factors affecting Trust}
In addition to the formal aspects, there were also various aspects relating to the content which promoted trust in the notification.
94 out of 377 (24.3\,\%) responses indicate the factual correctness and detailed explanation to be trust-promoting, and 56 (14.8\,\%) participants mentioned the same about the CheckGA tool.

For several participants, the underlying motivation of the sender was relevant.
76 out of 377 (20.2\,\%) participants considered it to be trust-promoting that there were no financial demands or profit-making intentions of the sender and that the notification did not contain any threat.
In contrast to this, 38 out of 252 (15.1\,\%) participants claimed that the sender's motivation was not clear, and 64 (25.4\,\%) participants perceived the notification as a threat, spam, or ad.

\paragraph{Verifiability Increases Trust}
While 11 out of 252 (4.4\,\%) participants stated that they generally do not trust information from unknown senders, in 119 out of 377 (31.6\,\%) responses, the possibility of verification was rated as trust-promoting. 
This includes that they could verify the sender, which some did by calling the provided number to ensure that the letter was sent by the claimed person.
For others, this includes verifying the facts through their own research or with the help of experts or acquaintances.

\subsection{Problem Solving and Support}
With our notification, we wanted to support the site owners.
Therefore, we asked to what extent the explanation in the notification and the \hidefornow{linked }self-service tool were helpful and whether the participants would like to receive notifications in the future.

\paragraph{Problem Solving} 
339 out of 437 (77.6\,\%) of the participants stated that they were able to understand the problem of missing IP Anonymization from the notification.
Many participants who had resolved the problem stated that they had done this without help (37.8\,\%), while 30.9\,\% reported that they asked their external service provider to resolve the problem.
13.0\,\% forwarded the issue to colleagues in the organization and 10.8\,\% resolved the problem themselves after getting help (other: 7.5\,\%, $N=362$).

\paragraph{Helpfulness of Self-Service Tool}

The CheckGA tool was rated as (very) helpful by the majority of participants (87.2\,\%; 266 out of 305).
This is in line with the fact that the tool has frequently been mentioned as a trust-building factor.
Another 51 respondents said they did not know the tool, and 86 participants stated they had not used it.

\paragraph{Future Notifications} Most owners (88.4\,\%) wish for future notifications about privacy issues on their website ($N = 448$).
The majority (84.8\,\%) preferred to be notified by email.
28.2\,\% preferred letter, 3.7\,\% a blog post, 3.2\,\% a call (1.7\,\% preferred something else, e.\,g., a service portal; $N = 401$; multiple answers possible).
30.5\,\% stated that they would be willing to pay for such notifications ($N=383$).





%% file: sections/07-discussion.tex
\section{Discussion}
\label{sec:discussion}
Our goal was to determine the factors that influence the success of a notification using both \hidefornow{technical }measurements and a survey of owners.
In this section, we review the most important results of both approaches and highlight where their results diverge.
We then compare our results with those of prior work.

\subsection{Observed Behavior}
Our experiment showed a high \hidefornow{overall }remediation rate, with 56.6\,\% of all notified operators remediating within two months, compared to 9.2\,\% of the control group.
Interestingly, the time required to remediate the issue is similar for most groups---a large portion of owners becomes compliant within the first seven days of message receipt, with a much smaller number following \hidefornow{suit }over the next weeks (cf. \hyperref[fig:survival-groups]{Figure~\ref*{fig:survival-groups}a}).
However, the spread of absolute remediation rates is high, ranging from 76.3\,\% to 33.9\,\%, indicating that the individual factors of the notification can have a significant influence on notification effectiveness.
We discuss these factors in greater detail here.

\paragraph{The Right Framing Can Have a Large Effect}
The framing of the message proved to be a major factor in notification success, with \gdpr{} and \fee{} significantly outperforming the baseline \privacy{} argument and an almost 20 percentage-point gap in survival between the extremes after the initial notification (all differences were statistically significant, both before the reminder and when considering the full timeframe).
Mentioning legal requirements and potential fines seems to increase the perceived severity of the issue and encourage the site owners to prioritize its resolution.

\paragraph{The Choice of Messenger Matters}
Our results show that \unilaw{} significantly outperforms the \unics{} group, achieving a remediation rate of 59.7\,\% compared to the 54\,\% of \unics{} ($p<0.05$), showing that the sender can make a significant difference in notification campaigns. The \private{} group shows only statistically insignificant differences to the other two groups, falling between them with a remediation rate of 56.1\,\%. We will return to this in \autoref{sec:discussion:comparison}.

\paragraph{Letters Increase Remediation, but at a Cost}
Sending a letter instead of an email while leaving the other factors unchanged had a highly significant impact ($p<0.0001$), increasing the remediation rate by between 3.9 and 17.9 percentage points (mean: 11.1).
However, this benefit should be weighed against the costs, which can be prohibitively expensive (we spent around 5000\,€ on domestic postage in total), especially for international notifications.

\paragraph{Reachability Remains a Challenge}
Despite manually collecting contact information from a source that is legally mandated to be correct, we did not reach all recipients.
The rate of undeliverable messages (3.5\,\% of letters, 5.8\,\% of emails for the first notification) was surprisingly high, considering the care we took in data collection.
This suggests that some of the websites were not well-maintained, which may also have contributed to the presence of misconfigurations.



\paragraph{Persistence Pays Off}
Although human error prevents us from conducting a comprehensive analysis of the effect of our reminders, some of the results, not affected by our mistake, show the reminder increasing the overall remediation rates.
The group that benefited most from the reminder was \unilaw{} -- \letter{} -- \fee{}, where 45.3\,\% of the websites that were still non-compliant one month after the first notification became compliant after our reminder (cf. \autoref{tab:survivalratesappx}).
Overall, 41.2\,\% of all notified owners remediated before the reminder, which was increased to 56.6\,\% by the reminder.

\paragraph{Strong Demand for Self-Service Tools}
Finally, many site owners benefited from our self-service tool CheckGA that helped them to understand the problem and to verify their remediation attempts.
46.9\,\% of all notified site owners used CheckGA at least once.
67.6\,\% of owners that had become compliant by the end of the study used CheckGA.
It was also discussed outside of our study, with several data protection experts and consultancies tweeting or writing articles about it, which may have contributed to the 3.1\,\% of the control group that also used the tool.


\subsection{Survey of Owners}
The results of the survey confirm the empirical results in many places.
However, in some places, we observed interesting discrepancies, which we will discuss here.

\paragraph{No Single Factor Consistently Increases Trust}
Our survey shows that even minor aspects of the notification, like the fact that the letters used the official university letterhead and contained contact information and a signature, were important aspects in evaluating the trustworthiness of the message.
Conversely, a minority of respondents actually listed many of these factors as \emph{reducing} their trust in the message, illustrating that no perfect solution for everyone exists.

\paragraph{Distrust of Unsolicited Messages is Rampant}
Despite our efforts to ensure that the message appears trustworthy to the recipient, we found that a significant number of recipients initially distrusted the message and contacted the designated contact person, sometimes in creative ways, to question the authenticity of the message and enquire after the motivation of sending it.
Recipients in the \private{} -- \letter{} group, in particular, were frequently puzzled by the willingness of a private individual to spend money on a stamp to inform them about a misconfiguration on their website and questioned the motives, asking about potential commercial interests or bad intentions.
However, this distrust did not necessarily translate into (in)action, as we will see next.

\paragraph{Perception–Action Relationship Inconclusive} 
We observed a discrepancy between the self-reported level of trust towards our messages (cf. \autoref{fig:likert}) and the actual remediation rates reported in \autoref{sec:results:notify}.
While messages from a private individual were rated as less trustworthy than those from a computer science group at a well-known university, they nevertheless resulted in similar remediation rates.
One possible explanation may be that the recipients questioned the motives of the sender and were fearing that the message was a prelude to legal action, thereby increasing the perceived pressure to act of the message.
However, as we did not collect data on this, it remains an open question for future research.

Similarly, while the self-reported trust into emails and letters are almost identical, the overall remediation rates show a spread of 11.2 percentage points (49.1\,\% for \mail{} compared to 60.3\,\% for \letter{} at the end of the study).
These differences may be explainable by a different \textit{base trust} for the two media, i.\,e., even if the letter did not seem particularly trustworthy, the fact that it was a letter already elevated it over the email.
This is supported anecdotally by messages from recipients, with several claiming that they would have ignored the notification if it had arrived as an email.
However, it may also be partially related to differences in message delivery success between \letter{} and \mail{} due to spam filters, the effect of which we cannot quantify.

\paragraph{System Owners Desire Support in Remediation}
204 owners (5.1\,\%) asked us for support.
We explained the concrete issue and, occasionally, provided code examples, which sometimes required multiple rounds of emails, but frequently resulted in successful remediation.
While such individual support is infeasible for larger notification campaigns, it illustrates the importance of providing a remediation guide and verification tool, which significantly reduced the time required to answer their questions and likely also reduced the total number.
Owners also reported passing on the notification to their web design agencies or data protection officers, with 44\,\% of owners reporting passing on the notification to a colleague or external contractor for resolution.
These external relationships may also have contributed to their lack of awareness of the problem, which we will discuss next.

\paragraph{Owners Unaware of Tracking on their Websites}
In the survey, 19.5\,\% of respondents reported that they were unaware that \gls{GA} was active on their website, indicating that they were not actually using the collected analytics data.
This impression is further supported by the fact that 36\,\% of remediating owners removed \gls{GA} from their website instead of adding IP Anonymization.
In emails and calls, 28 out of 764 website owners indicated that they were unaware that \gls{GA} was running on their website, and others reported that they had not looked at the analytics data in years or that it was added by their web designer without informing them.
This raises questions of liability and indicates that some fraction of \gls{GA} installations is unintentional or dormant, and thus can be removed without negative consequences for the site owners.

\subsection{Comparison with Prior Work}
\label{sec:discussion:comparison}
We will now discuss how our results agree with or diverge from previous work.

\paragraph{Problem Awareness}
Some prior studies asked recipients if they were aware of the issue(s) 
before receiving the notification, finding surprisingly high awareness.
Durumeric \emph{et al.}, notifying about the \emph{Heartbleed} issue, found that \emph{all} recipients were aware of the issue, and many had already made attempts to remediate it \cite{Durumeric2014}.
Li \emph{et al.} found that 46\,\% of recipients had been aware, and 16\,\% had already attempted to remediate \cite{Li2016Usenix}.
\c{C}etin \emph{et al.} similarly reported that 40\,\% of their recipients had previously attempted to remediate \cite{Cetin2017}.
Our results paint a similar picture: 58.9\,\% of owners knew about the IP anonymization feature–and 12.7\,\% even knew they were not using it. 
These studies indicate that awareness is a necessary, but not sufficient condition for remediation. 

\paragraph{Bounce Rates}
Only one prior study previously used manual address collection: Stock \emph{et al.} conducted a small-scale ($N=364$ over 10 groups) experiment \cite{Stock2018}. All 91 emails were delivered correctly, but 18 out of 67 letters (26.8\,\%) were returned because the recipient could not be found.
However, their sample was drawn from recipients who had not reacted to the automated notification, leading to self-selection effects.
With 3.5 to 5.8\,\%, our observed bounce rates were slightly higher for emails and lower for letters, respectively.
They remain much lower than many previous studies that used automated address collection, some of which observed bounce rates exceeding 50\,\% \cite{Cetin2017,Stock2016,Cetin2016}.
This shows that manual address collection can be translated into higher delivery success, but some sites will still remain unreachable.

To compare our address collection methodology with approaches from previous work, we analyzed all 4425 email addresses that we collected from the imprints. Three of the previous studies \cite{Stock2016,Cetin2017,Stock2018} sent notifications to common addresses such as \emph{\{info,abuse,security,host\-master,webmaster\}@domain.tld}. While some of the addresses we collected \emph{do} match one of the addresses used in previous work (41.0\,\% of addresses had the form \emph{info@domain.tld}, 0.8\,\% were \emph{\{webmaster,hostmaster\}@domain.tld}), a substantial proportion \emph{does not} match addresses used in previous work (21.1\,\% had a different address prefix, and 37.1\,\% of addresses listed an entirely different domain). Note, however, that we do not know whether the addresses used in previous work would have worked as well (while not being listed in the imprint). Thus, our reported data on the availability of these addresses only poses a lower bound.

\paragraph{Effects of Message Sender}
Three prior studies considered different senders: \c{C}etin \emph{et al.} sent messages posing as a private security researcher, a university, and a well-known anti-malware organization \cite{Cetin2016}.
Stock \emph{et al.} compared emails that appeared to come from a human from those appearing to come from an automated system \cite{Stock2018}, while Zeng \emph{et al.} collaborated with Google to send part of their messages via the Google Search Console, with the others being sent via Email from a UC Berkeley account \cite{Zeng2019}.
In all three studies, the differences between the different senders were small and, where this was reported, statistically insignificant.

At first glance, this conflicts with our results, which show \unilaw{} to be significantly more effective than \unics{}.
A possible interpretation is that name recognition does not make a difference (explaining why previous studies, even with the support of a well-known company like Google, did not observe significant differences).
Instead, recipients consider if the sender can and will plausibly \emph{impose consequences} for inaction.
They may believe that a computer science group is unlikely to pursue legal action, while a message citing legal regulations sent by a private individual or legal experts is a stronger incentive, as the sender poses a more plausible threat.
This would be in line with prior research into framing and incentives, which we will consider next.


\paragraph{Framing and Incentives}
Zeng \emph{et al.} compared different framings for issues such as outdated TLS configurations and misconfigured or expiring certificates, using either a user focus (explaining the impact on the user) or a technical focus (explaining the technical background) \cite{Zeng2019}.
Unlike our study, they did not observe statistically significant differences in remediation rates, which may be related to the fact that a main incentive (the fact that users may be blocked from accessing the website) was present in both framings.

Other studies used stronger incentives, like browser warnings \cite{Li2016WWW} or quarantining infected users and refusing them access to the Internet until they remediated \cite{Cetin2018,Cetin2019}.
Of particular note is the study by \c{C}etin \emph{et al.}, which compared email notifications with quarantine networks and found the latter to be more effective \cite{Cetin2019}.
These results indicate that providing direct incentives for remediation may be a promising avenue. Our study suggests that regulatory requirements and the associated fines can serve as such an incentive.

\paragraph{Recipients (Dis)trust}
Similar to our results, prior studies reported that recipients often mistrusted the notifications \cite{Stock2018,Cetin2018,Cetin2019,Zeng2019} and reached out for verification \cite{Cetin2018,Cetin2019}.
We found that while some factors were reported as improving trust,
the same factors were also decreasing trust for a minority of recipients.
Reliably establishing trust remains an unsolved challenge, especially due to the prevalence of fraudulent messages Internet users are faced with.


\paragraph{Support Tools}
Several prior studies reported that recipients asked for automated systems to assist in remediation \cite{Li2016Usenix,Li2016WWW,Zeng2019,Cetin2017}.
\c{C}etin \emph{et al.} conducted a study to evaluate the effect of providing a tool and found that providing or withholding it did not have a statistically significant effect on remediation \cite{Cetin2017}.
While we did not repeat this experiment, our results indicate that, regardless of the effect on remediation, providing a tool may have other benefits, such as simplifying support for recipients, potentially reducing the amount of support requests, and increasing trust.

\paragraph{Reminders}
Previous research on the effect of reminder messages has been inconclusive, with Stock \emph{et al.} finding a small effect \cite{Stock2016}, while Li \emph{et al.} found none \cite{Li2016Usenix}. In our case, 29.7\,\% of websites that were still non-compliant after the first message remediated after the reminder, with some groups showing over 40\,\% additional remediations  (cf. \autoref{tab:survivalratesappx}). Thus, reminders were obviously effective. The reasons for this discrepancy remain unclear. In our survey, some recipients named the reminders as a trust-promoting factor. Others had remediated incompletely, and completed the remediation after receiving the reminder. However, this does not explain why previous studies did not see similar results. More research is needed to understand the effectiveness of reminders.

\paragraph{Summary}
Our study confirms many of the results of previous studies: Gaining the recipients' trust is difficult, and providing them with automated systems to validate their fixes is perceived as helpful. 
We also once again observed that awareness does not necessarily lead to action, which indicates that it may be helpful to provide system operators with incentives for remediation, and potential negative consequences from inaction.
Such consequences can take the form of browser warnings that scare off customers \cite{Zeng2019}, denying end users access to the Internet \cite{Cetin2018,Cetin2019}, or potential fees for violating relevant legislation.
However, our results also call into question previous results by showing that the identity of the sender and the sending of reminders can have significant effects on overall remediation.
More research is needed to understand the interplay of these factors.

\subsection{Limitations}
\label{sec:limitations}

Regarding \emph{internal validity}, our study has four limitations. Firstly, there are two kinds of potential self-selection. The first kind affects the group assignment of those sites that either provide only an email \emph{or} a postal address in the imprint.
This is the case for 87 and 152 owners, respectively, i.\,e., about 6\,\% of non-\control{} recipients.
The second kind of self-selection affects the survey. Our participants can be assumed to have a higher trust in our messages since distrusting our messages makes them less likely to respond to our survey invitation.

Secondly, our compliance checker does not confirm cookie consent banners.
Thus, any tracking that takes place \emph{after} giving consent is not detected.
We could thus misdetect the introduction of a consent banner as removal of \gls{GA}.
We have found no indicator that this has happened during the two months of the study, but did not check all websites.

Thirdly, we sent incorrect reminders to part of the \letter{} -- \unilaw{} group, the effect of which we are unable to quantify.
However, the most important trends were already visible before the reminders were sent.
We also received indications that a low number of recipients received messages from more than one group (e.\,g., because more than one website was operated by the same web design agency but listed different owners in their imprints).
We are unable to quantify the potential effects this may have had on remediation due to observer effects based on the suspicion of being part of a study.

Fourthly, we used three different email servers, which may have led to different message delivery rates due to spam classification. 
As we did not control the mail servers, we could not subscribe to spam reporting services.
As in previous studies \cite{Cetin2016,Zeng2019}, this setup may have introduced an impossible-to-quantify bias.
We also found after the fact that the mail servers of \unilaw{} were not configured with \gls{SPF} and \gls{DKIM} records.
Nevertheless, we are hopeful that the different mail servers do not have a large effect on the deliverability of our notifications.
First of all, we observed similar bounce rates for the three different senders, and the rates at which CheckGA was accessed were actually \emph{highest} for the \unilaw{} group.
Secondly, we sent only relatively small numbers of mails, all of them with slightly different content, to individual mail servers. The 1337 recipient addresses are spread over 516 distinct second-level domains. For further clarification, we analyzed the diversity of affected mail servers a few months after the end of the study.
According to the results, the average number of addresses handled by individual mail providers is 2.5 (median: 1). Even commonly used mail providers like Google and Outlook.com received only 70 and 108 mails, respectively---and these mails were submitted over a period of five days using three different sending servers.

\emph{External validity} of our results is subject to two limitations.
Firstly, while we aimed to obtain a diversified set of websites, our sample is not representative for the overall population of websites in Germany.
Secondly, and more importantly, all of our observations relate to German site owners, i.\,e., it is unknown whether our insights apply to other countries with different legal regimes and cultures.
This limitation is a consequence of our compliance-focused approach. Compliance issues are rooted in local laws and have to be addressed specifically for every country. While compliance-based notifications appear to increase the pressure to act for German site owners, we cannot say anything about their effectiveness in other countries.
The effort of tailoring notifications on a per-country level may be higher, but this approach does have its advantages: better message comprehension and trust through name recognition of the involved organizations.
Thus, it may be a promising avenue for researchers to relay notifications through local partners (similar to \cite{Kuhrer2014,Li2016WWW,Li2016Usenix}) who can relate the issue to the respective local laws \cite{diop_coerce_2019}.

%% file: sections/09-conclusion.tex
\section{Conclusion}
\label{sec:conclusion}
\glsreset{GA}
Our study indicates that effective notification campaigns on the web are a matter of trust, framing, and support.
We reach this conclusion based on a covert experiment with 4594 website owners running 4754 websites that used \gls{GA} without IP Anonymization, i.\,e., failing to comply with current European data protection regulation.
Our notifications led to an overall remediation rate of 56.6\,\%, a significant increase compared to the 9.2\,\% of the control group.

In addition, a survey with 477 responses allowed us to identify a number of formal and content-related factors that influenced the recipients' trust in the notification.
We also collected first impressions of how website owners solved the problem and which support they benefited from, showing that there was a high demand for our self-service tool, but also for support via email or phone.
More research is needed into how this type of support can be standardized and scaled for larger notification campaigns. 



According to our results, reminding website owners about legal obligations can increase remediation rates by over 20 percentage points.
Thus, even senders without any authority to impose fines themselves can motivate site owners to remediate a misconfiguration.
Parties interested in running a notification campaign may be well-advised to consult with legal experts, not only to ensure the legality of their own notification but also to investigate if the topic of the notification can be framed as an issue of compliance.

Finally, we found that most website owners were unaware of their non-compliance before our notification, with 19.5\,\% of survey respondents not even being aware that their website was using \gls{GA}.
Further, 36\,\% of the remediating site owners chose to completely remove GA, and several website owners took their websites offline.
Thus, notification campaigns may also motivate website owners to disable unmaintained systems, including analytics tools whose data is never viewed or even complete sites that are obsolete to them, improving the privacy and security posture of the Web.

%% file: sections/10-acknowledgements.tex
\section*{Availability}
\urlstyle{same}
The code of our crawler and the CheckGA tool, the translated notices, the survey questions, and parts of the dataset that could be anonymized and its associated evaluation code can be found online \cite{SupplementaryMaterial}. The CheckGA tool (in German) can be accessed at \url{https://checkgoogleanalytics.psi.uni-bamberg.de/}.

\paragraph{Acknowledgements}
This work has been co-funded by the DFG as part of projects C.1 and D.5 within the RTG 2050 “Privacy and Trust for Mobile Users", and by the German BMBF and the Hessen State Ministry for Higher Education, Research and the Arts within their joint support of the National Research Center for Applied Cybersecurity ATHENE.



%% file: sections/11-appendix.tex

\vspace*{-2mm}
\section*{A1\ \ \ \ Significance Tests}
\label{appx:significance}
Tables \ref{tab:survivalrates} and \ref{tab:survivalratesappx} show survival rates after 35 (pre-reminder), 24 (reminder) and 55 (full time frame) days, respectively.
\autoref{tab:significance-tables} shows the corresponding significance levels, with p-values corrected for multiple tests with a single Holm-Bonferroni correction \cite{Holm1979} for all 45 significance tests.

%

\vspace*{-1mm}
\section*{A2\ \ \ \ Co-Owned Websites}
\label{app:grouping}
Websites that list the same contact information in their imprint are grouped as \emph{co-owned} and notified in a single message.
Aside from reducing the number of messages to be sent, this is done to model that these websites are related and may thus be remediated at the same time.
This grouping has to be considered a first approximation of the real operator structure, as websites may list different owners but be maintained by the same web design agency, which may take a notification as a reason to also check and repair other websites in their portfolio.
We received some indicators that this was the case, but lack a method to quantify the effect on the measurement.

\begin{table}[]
    \centering
    \caption{Survival rates in percent for pre- and post-reminder groups and at the end of the study (lower is better). Results marked with $\dagger$ may be impacted by human error, see \autoref{sec:results:notify:reminder}. Results are based on 1321 emails and 2644 letters.}
    \label{tab:survivalrates}
    \begin{tabular}{@{}llll@{}}
        \toprule
        Group & Pre-rem. & Post-rem. & End of study \\
        \midrule
        \mail{}    & $66.3 {\color{gray}\,\pm\,2.6}$ & $75.8 {\color{gray}\,\pm\,3.1}$ & $50.9 {\color{gray}\,\pm\,2.7}$ \\
        \letter{}  & $55.6 {\color{gray}\,\pm\,1.9}$ & $66.6 {\color{gray}\,\pm\,2.6}\dagger{}$ & $39.7 {\color{gray}\,\pm\,1.9}\dagger{}$ \\
        \midrule
        \private{} & $59.9 {\color{gray}\,\pm\,2.7}$ & $69.0 {\color{gray}\,\pm\,3.4}$ & $43.9 {\color{gray}\,\pm\,2.7}$ \\
        \unics{}   & $61.4 {\color{gray}\,\pm\,2.7}$ & $70.8 {\color{gray}\,\pm\,3.4}$ & $46.0 {\color{gray}\,\pm\,2.7}$ \\
        \unilaw{}  & $55.0 {\color{gray}\,\pm\,2.8}$ & $69.5 {\color{gray}\,\pm\,3.8}\dagger{}$ & $40.3 {\color{gray}\,\pm\,2.7}\dagger{}$ \\
        \midrule
        \privacy{} & $69.6 {\color{gray}\,\pm\,2.6}$ & $75.1 {\color{gray}\,\pm\,3.2}\dagger{}$ & $54.7 {\color{gray}\,\pm\,2.7}\dagger{}$ \\
        \gdpr{}    & $56.6 {\color{gray}\,\pm\,2.8}$ & $69.0 {\color{gray}\,\pm\,3.7}\dagger{}$ & $41.9 {\color{gray}\,\pm\,2.7}\dagger{}$ \\
        \fee{}     & $50.1 {\color{gray}\,\pm\,2.8}$ & $63.3 {\color{gray}\,\pm\,3.9}$ & $33.7 {\color{gray}\,\pm\,2.6}$ \\
        \midrule
        All notified & $58.8 {\color{gray}\,\pm\,1.6}$ & $70.3 {\color{gray}\,\pm\,2.0}\dagger{}$ & $43.4 {\color{gray}\,\pm\,1.6}\dagger{}$ \\
       \control{} & $93.0 {\color{gray}\,\pm\,2.4}$ & $97.6 {\color{gray}\,\pm\,1.7}$ & $90.8 {\color{gray}\,\pm\,2.6}$ \\
        \bottomrule
    \end{tabular}
\end{table}

\begin{table*}
\centering
\caption{Survival rates in percent for all groups. Results marked with $\dagger$ are potentially impacted by human error, see \autoref{sec:results:notify:reminder}. For remediation rates, subtract survival rates from 100.}
\label{tab:survivalratesappx}
\begin{tabular}{lllrrlll}
\toprule
Medium & Sender & Framing & Owners & Sites & Pre-rem. [\%] & Post-rem. [\%] & End of study [\%] \\
\midrule
\mail   & \private & \privacy & 146 & 163 & $79.8 {\color{gray}\,\pm\,7.5}$ & $80.7 {\color{gray}\,\pm\,8.7}$         &  $63.5 {\color{gray}\,\pm\,8.4}$ \\
        &          & \gdpr    & 149 & 153 & $63.8 {\color{gray}\,\pm\,8.2}$ & $77.8 {\color{gray}\,\pm\,10.1}$        &  $49.0 {\color{gray}\,\pm\,8.2}$ \\
        &          & \fee     & 148 & 159 & $64.3 {\color{gray}\,\pm\,8.3}$ & $74.1 {\color{gray}\,\pm\,10.3}$        &  $48.8 {\color{gray}\,\pm\,8.3}$ \\
        & \unics   & \privacy & 146 & 166 & $82.0 {\color{gray}\,\pm\,7.5}$ & $78.7 {\color{gray}\,\pm\,9.0}$         &  $66.1 {\color{gray}\,\pm\,8.3}$ \\
        &          & \gdpr    & 149 & 152 & $62.4 {\color{gray}\,\pm\,8.3}$ & $74.7 {\color{gray}\,\pm\,10.8}$        &  $47.0 {\color{gray}\,\pm\,8.2}$ \\
        &          & \fee     & 145 & 147 & $61.0 {\color{gray}\,\pm\,8.5}$ & $63.4 {\color{gray}\,\pm\,11.4}$        &  $39.3 {\color{gray}\,\pm\,7.9}$ \\
        & \unilaw  & \privacy & 147 & 149 & $65.6 {\color{gray}\,\pm\,8.3}$ & $88.1 {\color{gray}\,\pm\,9.1}$         &  $55.6 {\color{gray}\,\pm\,8.4}$ \\
        &          & \gdpr    & 144 & 147 & $65.6 {\color{gray}\,\pm\,8.3}$ & $78.8 {\color{gray}\,\pm\,10.7}$        &  $53.1 {\color{gray}\,\pm\,8.5}$ \\
        &          & \fee     & 147 & 149 & $52.3 {\color{gray}\,\pm\,8.3}$ & $67.6 {\color{gray}\,\pm\,12.5}$        &  $35.4 {\color{gray}\,\pm\,7.7}$ \\
\letter & \private & \privacy & 294 & 308 & $69.2 {\color{gray}\,\pm\,5.6}$ & $70.6 {\color{gray}\,\pm\,7.1}$         &  $52.9 {\color{gray}\,\pm\,5.9}$ \\
        &          & \gdpr    & 294 & 304 & $50.5 {\color{gray}\,\pm\,5.8}$ & $60.9 {\color{gray}\,\pm\,8.8}$         &  $33.0 {\color{gray}\,\pm\,5.4}$ \\
        &          & \fee     & 292 & 298 & $48.4 {\color{gray}\,\pm\,5.8}$ & $59.0 {\color{gray}\,\pm\,8.9}$         &  $30.9 {\color{gray}\,\pm\,5.4}$ \\
        & \unics   & \privacy & 294 & 302 & $68.5 {\color{gray}\,\pm\,5.6}$ & $76.7 {\color{gray}\,\pm\,6.7}$         &  $55.8 {\color{gray}\,\pm\,5.9}$ \\
        &          & \gdpr    & 292 & 305 & $54.6 {\color{gray}\,\pm\,5.9}$ & $65.0 {\color{gray}\,\pm\,8.7}$         &  $39.8 {\color{gray}\,\pm\,5.6}$ \\
        &          & \fee     & 293 & 303 & $51.9 {\color{gray}\,\pm\,5.8}$ & $64.4 {\color{gray}\,\pm\,8.5}$         &  $35.4 {\color{gray}\,\pm\,5.5}$ \\
        & \unilaw  & \privacy & 293 & 293 & $62.5 {\color{gray}\,\pm\,5.8}$ & $70.4 {\color{gray}\,\pm\,7.5}\dagger$  &  $44.7 {\color{gray}\,\pm\,5.8}\dagger$ \\
        &          & \gdpr    & 288 & 294 & $55.6 {\color{gray}\,\pm\,5.9}$ & $68.5 {\color{gray}\,\pm\,8.2}\dagger$  &  $41.3 {\color{gray}\,\pm\,5.7}\dagger$ \\
        &          & \fee     & 293 & 304 & $39.4 {\color{gray}\,\pm\,5.6}$ & $54.7 {\color{gray}\,\pm\,10.0}$        &  $23.7 {\color{gray}\,\pm\,5.0}$ \\
        \midrule
        \multicolumn{3}{l}{All notified} & 3954 & 4096 & $58.8 {\color{gray}\,\pm\,1.6}$ & $70.3 {\color{gray}\,\pm\,2.0}\dagger$ & $43.4 {\color{gray}\,\pm\,1.6}\dagger$ \\
        \multicolumn{3}{l}{\control{}} & 585 & 600 & $93.0 {\color{gray}\,\pm\,2.4}$ & $97.6 {\color{gray}\,\pm\,1.7}$ & $90.8 {\color{gray}\,\pm\,2.6}$ \\
        \bottomrule
\end{tabular}
\end{table*}

\begin{table*}
    \caption{Significance levels for comparison of survival rates for different senders, framings, and media at different points in time.}
    \label{tab:significance-tables}
    \includegraphics[width=1\textwidth]{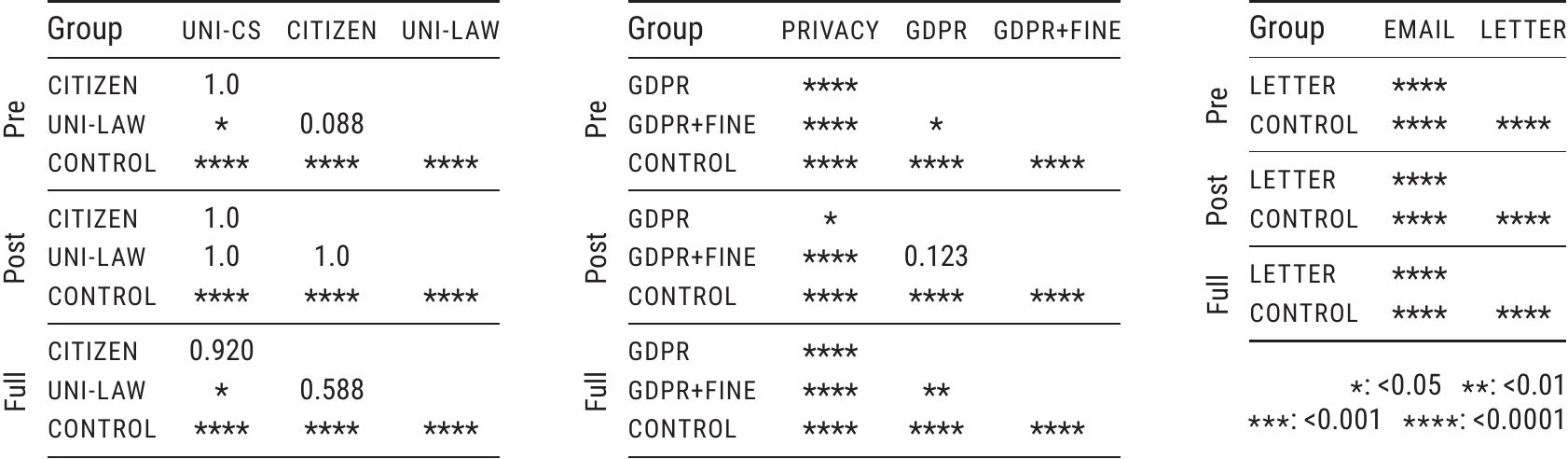}
\end{table*}

One might intuitively assume that, if an owner is notified about its non-compliance for more than one website, it would ensure that all websites are made compliant (or remain non-compliant if they choose to ignore the notification).
In fact, websites run by 77 out of 88 owners with more than one website (87.5\%) were either completely compliant or non-compliant at the end of the study timeframe (40 compliant, 37 non-compliant, counting groups assigned to the control group).
Of the 40 compliant owners, 34 (85\%) made all websites compliant within a timeframe of at most 2 days between first and last remediation, with the remaining six groups taking between 4 and 35 days to remediate the other websites.


\lstdefinelanguage{JavaScript}{
  keywords={typeof, new, true, false, catch, function, return, null, catch, switch, var, if, in, while, do, else, case, break},
  keywordstyle=\color{blue}\bfseries,
  ndkeywords={class, export, boolean, throw, implements, import, this},
  ndkeywordstyle=\color{darkgray}\bfseries,
  identifierstyle=\color{black},
  sensitive=false,
  comment=[l]{//},
  morecomment=[s]{/*}{*/},
  commentstyle=\color{purple}\ttfamily,
  stringstyle=\color{red}\ttfamily,
  morestring=[b]',
  morestring=[b]"
}

\lstset{
   language=JavaScript,
   backgroundcolor=\color{white},
   extendedchars=true,
   basicstyle=\ttfamily,
   showstringspaces=false,
   showspaces=false,
   numbers=left,
   numberstyle=\footnotesize,
   numbersep=9pt,
   tabsize=2,
   breaklines=true,
   showtabs=false,
   captionpos=t
}
\begin{lstlisting}[caption=Examples of erroneous IP Anonymization configurations for Google Analytics using analytics.js, float=*, label=lst:analyticsjs]
(function(i,s,o,g,r,a,m){i['GoogleAnalyticsObject']=r;i[r]=i[r]||function(){
(i[r].q=i[r].q||[]).push(arguments)},i[r].l=1*new Date();a=s.createElement(o),
m=s.getElementsByTagName(o)[0];a.async=1;a.src=g;m.parentNode.insertBefore(a,m)
})(window,document,'script','https://www.google-analytics.com/analytics.js','ga');
ga('set', 'anonymizeIp', true);     // Error: Must be done after configuring the ID
ga('create', 'UA-XXXXX-Y', 'auto'); // Configure the tracking ID
ga('set', 'anonymizeIP', true);     // Error: Must be spelled 'anonymizeIp'
ga('send', 'pageview');             // Send the pageview
ga('set', 'anonymizeIp', true);     // Error: Must be done before sending the pageview
\end{lstlisting}

\vspace*{-2mm}
\section*{A3\ \ \ \ Google Analytics Misconfigurations}
\vspace*{-1mm}
\label{app:misconfiguration}
When activating IP Anonymization (AIP) for Google Analytics, operators can encounter several pitfalls.
First, AIP must be activated explicitly.
Google implements AIP since May 2010.
Operators who included GA earlier must be aware of this addition and change their website.
Secondly, how to enable AIP depends on how GA is included, e.\,g., for inclusion via Google Tag Manager, the option must be set in Google’s web interface, while adding the Analytics library via a <script> tag requires additional JavaScript code to enable the option.
Thirdly, there are several versions (analytics.js and ga.js) of the Analytics library, which require different approaches to activate AIP.
There are also loaders such as gtag.js, which load these libraries, adding more variety in the approaches.
Fourthly, the option to enable AIP is case-sensitive and spelled “anonymizeIp”, except in gtag.js, where the option is called ``anonymize\_ip''.
Note the lowercase “p” in “Ip”, which is likely to be misspelled “IP.”
Misspelling does not raise an error but silently ignores the option.
Fifthly, the “anonymizeIp” option must be set 
after configuring the tracking ID but before any requests to Google are issued.
Again, there is no warning that non-anonymized requests are issued when setting this option too early or too late.
Finally, an operator can define several GA trackers on a single page.
For each defined tracker, AIP has to be enabled separately.
An example of different misconfigurations is shown in \autoref{lst:analyticsjs}.